\documentclass[preprint,aps,nofootinbib]{revtex4}

\usepackage{graphicx}
\usepackage{latexsym}
\usepackage{amssymb}
\usepackage{epsfig}

\usepackage{bm}

\newcommand{\be}{\begin{equation}}
\newcommand{\ee}{\end{equation}}
\newcommand{\beqq}{\setlength\arraycolsep{2pt}\begin{eqnarray}}
\newcommand{\eeqq}{\vspace{0cm} \end{eqnarray}}
\newcommand{\bea}{\begin{eqnarray}}
\newcommand{\eea}{\end{eqnarray}}

\newcommand{\bn}{\begin{eqnarray}}
\newcommand{\en}{\end{eqnarray}}

\newcommand{\p}{\partial}

\newcommand{\no}{\noindent}

\def\bea{\begin{eqnarray}}
\def\eea{\end{eqnarray}}

\newcommand{\beq}{\begin{eqnarray}}
\newcommand{\eeq}{\end{eqnarray}}

\begin{document}

\title{Cosmological bounds on open FLRW solutions of massive gravity}

\author{S. H. Pereira} \email{shpereira@gmail.com}
\author{E. L. Mendon\c ca} \email{elias@gmail.com}
\author{A. Pinho S. S.} \email{alexandre.pinho510@gmail.com}

\affiliation{Faculdade de Engenharia de Guaratinguet\'a \\ UNESP - Univ. Estadual Paulista ``J\'ulio de Mesquita Filho''\\ Departamento de F\'isica e Qu\'imica\\ Av. Dr. Ariberto Pereira da Cunha 333 - Pedregulho\\
12516-410 -- Guaratinguet\'a, SP, Brazil}

\author{J. F. Jesus} \email{jfjesus@itapeva.unesp.br}
\affiliation{UNESP - Univ. Estadual Paulista ``J\'{u}lio de Mesquita Filho''\\ C\^ampus Experimental de Itapeva - R. Geraldo Alckmin, 519\\ Vila N. Sra. de F\'atima, 18409-010 -- Itapeva, SP, Brazil}



\begin{abstract}
In this work we have analysed some cosmological bounds concerning an open FLRW solution of massive gravity. The constraints with recent observational $H(z)$ data were found and the best fit values for the cosmological parameters are in agreement with the $\Lambda$CDM model, and also point to a nearly open spatial curvature, as expected from the model. The graviton mass dependence with the constant parameters $\alpha_3$ and $\alpha_4$, related to the additional lagrangians terms of the model, are also analysed, and we have obtained a strong dependence with such parameters, although the condition $m_g\simeq H_0^{-1}$ seems dominant for a long range of the parameters $\alpha_3$ and $\alpha_4$. 

\end{abstract}

\maketitle


\section{Introduction}

Current observations of Supernovae type Ia (SNIa) \cite{SN,union2}, Cosmic
Microwave Background (CMB) radiation \cite{WMAP,planck} and Hubble parameter data \cite{farooq,sharov} indicate an accelerated
expansion of the universe, being the $\Lambda$CDM model the best model to
fit the observational data. The $\Lambda$ term corresponds to a
cosmological constant energy density which is plagued with several
fundamental issues \cite{CC}, which has motivated the search for
alternatives models of gravity that could explain the observations. 

Massive gravity theories \cite{fierz,deser,hassan1,hassan2,hassan3,rham,hinterRMP,massiveG,volkov,koba,gumru1,gumru2,rham2} are old
candidates to explain the accelerated expansion of the universe, since
that the graviton mass could perfectly induce and mimic a cosmological
constant term. However, such kinds of theories were considered for
long time as being unsuitable due to the appearing of Boulware-Deser
(BD) ghosts \cite{deser}. Recently it was discovered a nonlinear
massive gravity theory that was first shown to be BD ghost free by Hassan et al. \cite{hassan1,hassan2}, and also in the Stuckelberg formulation in \cite{hassan3}. Then such theory was also developed by de Rham et al.\cite{rham}, sometimes called dRGT (de Rham-Gabadadze-Tolley) model
(see \cite{hinterRMP} for a review), bringing
back the cosmological interest for such theories
\cite{massiveG}. Self-accelerating cosmologies with ghost-free massive
gravitons have been studied thereafter \cite{volkov,koba,gumru1,gumru2,rham2}. Nowadays there is a general agreement that in dRGT models which are defined with a flat reference metric, isotropic flat and closed FLRW cosmologies do not exist, even at the background level. Nevertheless isotropic open cosmologies exist as classical solutions but have unstable perturbations. In the case of a non-flat reference metric a ghost free theory exist for all types of background cosmologies \cite{hassan2}, though the perturbations are still unstable in isotropic cases \cite{gumru2}.

Although the theoretical aspects of massive gravity has been severely studied in the last years, the cosmological constraints with observational data does not. In \cite{saridakis} has been investigated the cosmological behavior in the quasi-Dilaton nonlinear massive gravity, and the parameters of the theory has been constrained with observational data from SNIa, Baryon Acoustic Oscillations (BAO) and CMB.

In order to study some cosmological bounds concerning the free parameters of the theory, we must have a massive gravity theory that has a FLRW limit well established for the reference metric. As showed in \cite{gumru2} the perturbations in unisotropic FLRW are stable, thus if we suppose that the unisotropies that cure the instability of the model do not greatly affect the cosmic evolution, we can use the results involving isotropic solutions as a first approximation to the stable unisotropic case. For this case the results of \cite{gumru2} are a good starting point to study some bounds in the massive gravity theory compared to recent observational data. In Section II we present the general massive gravity theory and the cosmological equations in Section III. In Section IV we present the constraints from $H(z)$ data and in Section V some bounds on the graviton mass are presented. We conclude in Section VI.

\section{Massive gravity theory}

Our starting point for the cosmological analysis is the massive theory for gravity proposed in \cite{rham}. The nonlinear action besides a functional of the physical metric $g_{\mu\nu}(x)$ include four spurious scalar fields $\phi^{a}(x)$ with $a=0,1,2,3$ called the St\"{u}ckelberg fields. They are introduced in order to make the action manifestly invariant under diffeomorphism, see for example \cite{dubowski}. Let us start by observing that these scalar fields are related to the physical metric and enter into the action as follows:

\be g_{\mu\nu}=f_{\mu\nu}+H_{\mu\nu}\label{metrica}\ee

\no where it is defined the fiducial metric $f_{\mu\nu}$ which is written in terms of the St\"{u}ckelberg fields:

\be f_{\mu\nu} \equiv \tilde{f}_{ab}(\phi^c)\p_{\mu}\phi^a\p_{\nu}\phi^b.\ee

\no Usually $\tilde{f}_{ab}$ is called the reference metric and for the purpose of this paper, one can use $\tilde{f}_{ab}=\eta_{ab}=(-,+,+,+)$ once the scheme proposed by dRGT respects Poincar\'e symmetry. Hence the fiducial metric in (\ref{metrica}) is nothing but the Minkowski metric in the coordinate system defined by the St\"{u}ckelberg fields. So we have automatically defined the covariant tensor $H_{\mu\nu}$ which propagates on Minkowski space and  the action is then a functional of the fiducial metric and the physical metric $g_{\mu\nu}$. 

The covariant action for massive general relativity that we are going to work with, can be written as:

\be S=M_{Pl}^2\int d^4x\,\,\sqrt{-g}\left\lbrack \frac{R}{2}+m^2{\cal U}(g,H)\right\rbrack\label{complete}\ee

\no where ${\cal U}$ is a potential without derivatives in the interaction terms between $H_{\mu\nu}$ and $g_{\mu\nu}$ that gives mass to the spin-2 mode described by the Einstein-Hilbert term. As observed in \cite{rham} a necessary condition for the theory (\ref{complete}) to be free of the Bouware-Deser ghost in the decoupling limit is that $\sqrt{-g}\,\,{\cal U}(g,H)$ be a total derivative. The most general covariant mass term which respects this condition is composed by\footnote{In \cite{rham2} a most general formulation is presented, where the parameters $\alpha_i$ are assumed to be dependent on the St\"{u}ckelberg fields.}:

\be \int d^4x\,\,\sqrt{-g}\,\,{\cal U}(g,H)=\int d^4x\sqrt{-g}\,\,(\alpha_2{\cal{L}}_2+\alpha_3{\cal{L}}_3+\alpha_4{\cal{L}}_4)\label{three}\ee
 where $\alpha_i$ are constants and the three lagrangians in (\ref{three}) are written as:

\bea {\cal L}_2&=&\frac{1}{2}([{\cal K}]^2-[{\cal K}^2])\\
{\cal L}_3&=&\frac{1}{6}([{\cal K}]^3-3[{\cal K}][{\cal K}^2]+2[{\cal K}^3])\\
{\cal L}_4&=&\frac{1}{24}([{\cal K}]^4-6[{\cal K}]^2[{\cal K}^2]+3[{\cal K}^2]^2+8[{\cal K}][{\cal K}^3]-6[{\cal K}^4]),\eea

\no where one has defined the tensor ${\cal K}_{\nu}^{\mu}=\delta^{\mu}_{\nu}-\sqrt{\p^{\mu}\phi^a\p_{\nu}\phi^b\eta_{ab}}$. In general there are other polynomial terms in ${\cal K}$ and the procedure to generate it can be found in \cite{hassan2,rham}, however it has been shown that all terms after the quartic order vanishes (see also \cite{matas,ref05}). Notice also that we have maintained the constant term $\alpha_2$ in order to see its consequences on the evolution equations. The case $\alpha_2=0$ is pathological since in this case the linearised theory and the non-linear one have different number of propagating modes.  

\section{Cosmology of massive gravity}

Let us begin by considering an open $(K<0)$, homogeneous and isotropic FRW universe for the physical metric:
\be g_{\mu\nu}dx^{\mu}dx^{\nu}=-N(t)^2dt^2+a(t)^2\left\lbrack dx_idx^i+\frac{K(x_idx^i)^2}{1-Kx_ix^i}\right\rbrack,\ee
where, $\mu,\nu=0,1,2,3$ and $i,j=1,2,3$, with $x^0=t$, $x^1=x$, $x^2=y$, $x^3=z$. Adopting the same {\it ansatz} for the St\"{u}ckelberg fields used in \cite{gumru2}, i.e:

\be \phi^0= f(t)\sqrt{1-Kx_ix^i}\quad; \quad \phi^i=\sqrt{-K}f(t)x^i,\label{stu}\ee

\no after plugging back the metric and (\ref{stu}) in (\ref{complete}), one obtains the following Lagrangian \cite{gumru2} for $a(t)$ and $f(t)$, where overdot will denote the time derivative:
\be
{\cal L}_g={1\over 8\pi G}\bigg[3K N(t) a(t) -
  {3\dot{a}(t)^2a(t)\over N(t)} + m_g^2(\alpha_2 L_2+\alpha_3L_3+\alpha_4L_4)\bigg]\,,\label{Sg}
\ee
where
\beqq
L_2&=&(3a(t)^2-3a(t)\sqrt{-K}f(t))(2N(t) a(t)-\dot{f}(t) a(t)-N(t)\sqrt{-K}f(t))\,,\nonumber\\
L_3&=&(a(t)-\sqrt{-K}f(t))^2(4N(t) a(t)-3\dot{f}(t)a(t)-N\sqrt{-K}f(t))\,,\nonumber\\
L_4&=&(a(t)-\sqrt{-K}f(t))^3(N(t)-\dot{f}(t))\,.
\eeqq
The matter content is assumed to be of the form
$T^\mu_\nu=\mathrm{diag}[-\rho_m(t),\,p_m(t),\, p_m(t),\,p_m(t)]$. Taking the Euler-Lagrange equation of ${\cal L}_g$ with respect to $f$ leads to
\be
(\dot{a}(t)-\sqrt{-K}N(t))[\alpha_2(3-2C)+\alpha_3(3-4C+C^2)+\alpha_4(1-2C+C^2)]=0\,,
\ee
where $C=f(t)\sqrt{-K}/a(t)$. The two interesting solutions are\footnote{The case $\dot{a}(t)=\sqrt{-K}N(t)$ just reproduces a constant scale factor when we take $N\to 1$ in order to recover the FLRW metric.}:
\be
C_\pm ={\bigg(\alpha_2+2\alpha_3+\alpha_4\pm \sqrt{\Delta}\bigg)\over (\alpha_3+\alpha_4)}\,,\label{f}
\ee
where $\Delta = \alpha_2(\alpha_2+\alpha_3-\alpha_4)+\alpha_3^2$. We take Euler-Lagrange equation of (\ref{Sg}) with respect to $N$
and use (\ref{f}) to obtain the Friedmann equation:
\be
{\dot{a}(t)^2\over N(t)^2 a(t)^2}+{K\over a(t)^2}={8\pi G\over
  3}\rho_m+{\Lambda_\pm\over3}\,,\label{friedmann}
\ee
where $\Lambda_\pm=m_g^2\beta_\pm$ and
\be
\beta_\pm = -{1\over (\alpha_3+\alpha_4)^2}\bigg[{2\alpha_2^3}+{3\alpha_2^2}(\alpha_3-\alpha_4\pm \sqrt{\Delta}) +{3\alpha_2}(\alpha_3-\alpha_4)(\alpha_3\pm \sqrt{\Delta}) - (\alpha_3\pm \sqrt{\Delta})^2(-2\alpha_3\pm \sqrt{\Delta})\bigg]\,\label{Lambda}
\ee
is a dimensionless parameter depending only on the constants
$\alpha_2$, $\alpha_3$ and $\alpha_4$. In this equation one can
recognize $\Lambda_\pm$ as being the energy density of massive gravity ($\rho_g$). Equally, it is expected to find the pressure term ($p_g$) in the second Friedmann equation.

Such equation can be obtained by combining Eq. (\ref{friedmann}) with
the variation of Eq. (\ref{Sg}) with respect to $a(t)$. Namely, we
subtract Eq. (\ref{friedmann}) from the Euler-Lagrange equation of $a(t)$ in order to get
\be
{-{2\dot{H}(t)\over N(t)}+{2 K\over a(t)^2}={8\pi G\over 3}(\rho_m +p_m)}\,\label{friedmann2}
\ee
where $H={\dot{a}(t)\over N(t) a(t)}$. Notice that the r.h.s of the
above equation contains only the contribution from the matter part,
which indicates that the graviton mass contribution satisfies an
equation of state of the form $p_g=-\rho_g$, exactly as a vacuum behaviour. This same result was observed in \cite{gumru2}.

Another combination is possible in order to get a direct relation
among $\ddot{a}(t)$, pressure and energy density of matter and
graviton. At this time we eliminate ${\dot{a}(t)^2\over N(t)^2 a(t)^2}$ of the expression from variation of Eq. (\ref{Sg}) with respect to $a(t)$. Thus, it is possible to get
\be
{\ddot{a}(t)\over N(t) a(t)}={\Lambda_\pm\over3}-{4\pi G\over 3}(\rho_m +3 p_m).\label{ddota}
\ee 
With such equation it is much easier to analyse the universe
acceleration. We conclude that an accelerated expansion occurs when ${\Lambda_\pm}>4\pi G(\rho_m +3 p_m)$. 

It is also easy to see that $\Lambda_\pm$ acts exactly like an
effective cosmological constant in Eq. (\ref{friedmann}). In both
Friedmann equations, (\ref{friedmann}), (\ref{friedmann2}) and also
(\ref{ddota}), there should be set $N=1$ in order to reproduce a
cosmological scenario. In order to reproduce a positive cosmological
constant (which leads to an accelerating universe), we must have $\Lambda_\pm >0$, which implies $\beta_\pm >0$.  

From now on we will assume $\alpha_2=1$ according to the original dRGT theory \cite{rham}. The Friedmann equation (\ref{friedmann}) can be rewritten in terms of
the present critical energy density $\rho_{c}=3H_0^2/8\pi G$,
\be
H(t)^2=H_0^2\bigg({\rho_m\over
  \rho_{c}}\bigg)+{m_g^2\beta_\pm}-{K\over a^2}\,,\label{H1}
\ee
where $H(t)=\dot{a}/a$ is the Hubble parameter and $H_0\simeq 70$ km/s/Mpc is its
present day value. Writing $\rho_m=\rho_{m0}(a/a_0)^{-3}$, where
$\rho_{m0}$ is the present day value for the matter energy density, and
introducing the density parameters 
\be
\Omega_{m}\equiv{\rho_{m0}\over \rho_{c}}\,,\hspace{1cm}
\Omega_{g}\equiv \beta_\pm{m_g^2\over H_0^2}\,,\hspace{1cm}
\Omega_{K}\equiv -{K\over a_0^2 H_0^2}\,,\label{omegas}
\ee
the Friedmann equation (\ref{H1}) can be expressed as
\be
H(t)^2=H_0^2\bigg[\Omega_{m}\bigg({a_0\over a}\bigg)^3
  +\Omega_{K}\bigg({a_0\over a}\bigg)^2 + \Omega_{g}\bigg]\,,
\ee  
or in terms of the redshift parameter, defined by $1+z \equiv a_0/a$,
\be
H(z)^2=H_0^2\bigg[\Omega_{m}(1+z)^3
  +(1-\Omega_{m}-\Omega_g)(1+z)^2 + \Omega_{g}\bigg]\,,\label{Hz}
\ee  
where we have used the Friedmann constraint
\be
1=\Omega_{m}+\Omega_{K}+\Omega_g\,,\label{fc}
\ee
that follows from (\ref{H1}) and (\ref{omegas}). Observational data can be used to constrain the values of such parameters and this will be done in the next section.

\section{Constraints from Observational $H(z)$ Data}

Observational $H(z)$ data provide one of the most straightforward and model independent tests of cosmological models, as $H(z)$ data estimation relies on astrophysical rather than cosmological assumptions. In this work, we use the data compilation of $H(z)$ from Sharov and Vorontsova \cite{sharov}, which is, currently, the most complete compilation, with 34 measurements.

From these data, we perform a $\chi^2$-statistics, generating the $\chi^2_H$ function of free parameters:
\begin{equation}
 \chi^2_H=\sum_{i=1}^{34}\left[\frac{H_0E(z_i,\Omega_m,\Omega_g)-H_i}{\sigma_{Hi}}\right]^2
\end{equation}
where $E(z)\equiv\frac{H(z)}{H_0}$ and $H(z)$ is obtained by Eq. (\ref{Hz}).

As the function to be fitted, $H(z)=H_0E(z)$, is linear on the Hubble constant, $H_0$, we may analytically project over $H_0$, yielding $\tilde{\chi}^2_H$:
\begin{equation}
\tilde{\chi}^2_{H}=C-\frac{B^2}{A}
\end{equation}
where $A\equiv\sum_{i=1}^{n}\frac{E_i^2}{\sigma_{Hi}^2}$, $B\equiv\sum_{i=1}^n\frac{E_iH_i}{\sigma_{Hi}^2}$, $C\equiv\sum_{i=1}^n\frac{H_i^2}{\sigma_{Hi}^2}$ and $E_i\equiv\frac{H(z_i)}{H_0}$.

\vspace{0.5cm}

\begin{figure}[t]
\centerline{\epsfig{figure=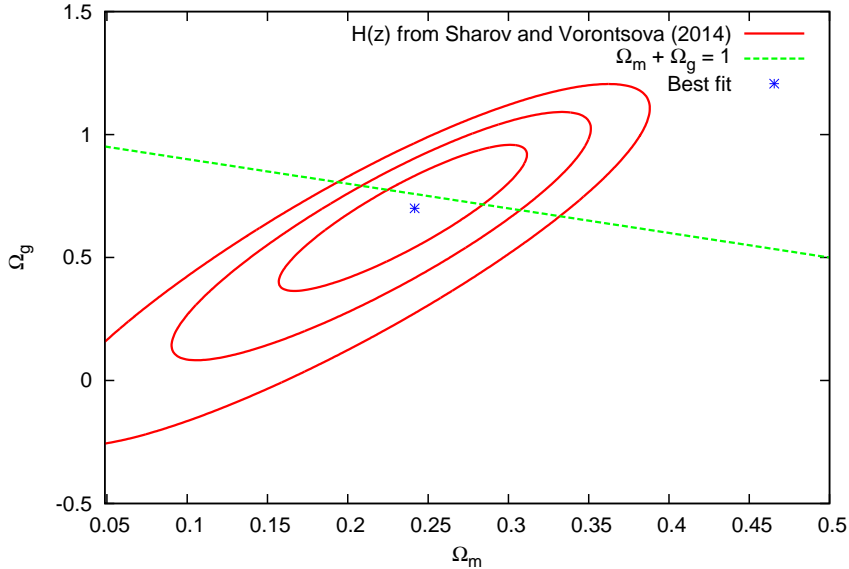,width=0.70\linewidth,angle=0}}
\caption{{\bf Solid lines:} Statistical confidence contours of massive gravity from $H(z)$ data. The regions correspond to 68.3\%, 95.4\% and 99.7\% c.l. {\bf Dashed line:} Flatness limit, where $\Omega_m+\Omega_g=1$. Points above this line are not considered on the statistical analysis. {\bf Star point:} Best fit, corresponding to $(\Omega_m,\Omega_g)=(0.242,0.703)$, which leads to $\Omega_K=0.055$. More details on the text.}
\label{contours}
\end{figure}

The result of such analysis can be seen on Figure \ref{contours}. As can be seen, the results from $H(z)$ data alone yield nice constraints on the plane $\Omega_m$ - $\Omega_g$. The flatness limit, which corresponds to $\Omega_m+\Omega_g=1$, can be seen as an straight line on this plane (dashed line on Fig. \ref{contours}). Points on and above this line were not considered, as they correspond to non-open models. One may see that the best fit relies right below this line, indicating that $H(z)$ data alone favour a slightly open Universe.

Furthermore, we have considered the prior $\Omega_m\geq\Omega_b$, with the baryon density parameter, $\Omega_b$, estimated by Planck and WMAP: $\Omega_b=0.049$ \cite{planck}, a value which is in agreement with Big Bang Nucleosynthesis (BBN), as shown on Ref. \cite{pdg}. As a result of this prior, the 3$\sigma$ c.l. contour alone is cut for low matter density parameter, as we may see on Fig. \ref{contours}.

The minimum $\chi^2$ was $\chi^2_{min}=16.727$, yielding a $\chi^2$ per degree of freedom $\chi^2_\nu=0.523$. The best fit parameters were $\Omega_m=0.242^{+0.041+0.065+0.090}_{-0.085-0.15\,\,\,-0.19}$,
 $\Omega_g=0.703^{+0.069+0.085+0.10}_{-0.34\,\,\,-0.62\,\,\,-0.96}$, for 68.3\%, 95.4\% and 99.7\% c.l., respectively, in the joint analysis.

As expected, this result is in agreement with $\Lambda$CDM
constraints, as this model mimics the concordance model. Moreover, the best fit values of $\Omega_m$ and $\Omega_g$ leads to $\Omega_K=0.055$, which corresponds to a negative value of $K$, as expected for this model. Sharov and Vorontsova \cite{sharov} have found, for
$\Lambda$CDM: $\Omega_m=0.276^{+0.009}_{-0.008}$,
$\Omega_\Lambda=0.769\pm0.029$, for 1$\sigma$ c.l., where they have
combined $H(z)$ with SN Ia and BAO data. Given the uncertainties on
massive gravity parameters above, the results are in good agreement,
even considering the open Universe restriction for massive gravity,
while $\Lambda$CDM has no restriction on curvature.

\section{Bounds on the graviton mass}

\begin{figure}[t]
\begin{center}
\epsfig{file=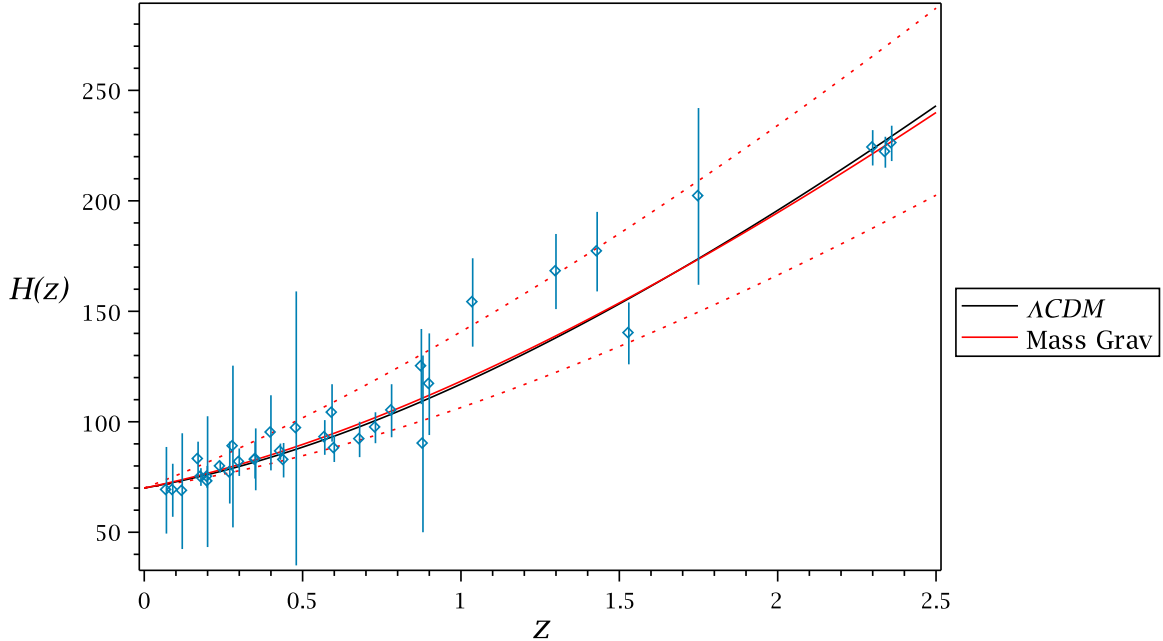, scale=0.8}
\caption{Plot of $H(z) \times z$ for the best fit values of Massive Gravity theory (red line, from this paper, $H(z)$ data only) with $\pm 1\sigma$ (red dotted line). The $\Lambda$CDM model is also represented (black line, best fit from Ref. \cite{sharov}, $H(z)$+BAO+SNs). The points with error bars are the Sharov and Vorontsova observational data \cite{sharov}.}\label{fig1}
\end{center}
\end{figure}

Having obtained the best fit values for the parameters, we show in Fig. \ref{fig1} the plot of Massive Gravity theory (red line) with $\pm 1 \sigma$ limit (red dotted line). The $\Lambda$CDM according to best fit data of Sharov and Vorontsova \cite{sharov} are also represented (black line). 

\begin{figure}[h]
\begin{center}
\epsfig{file=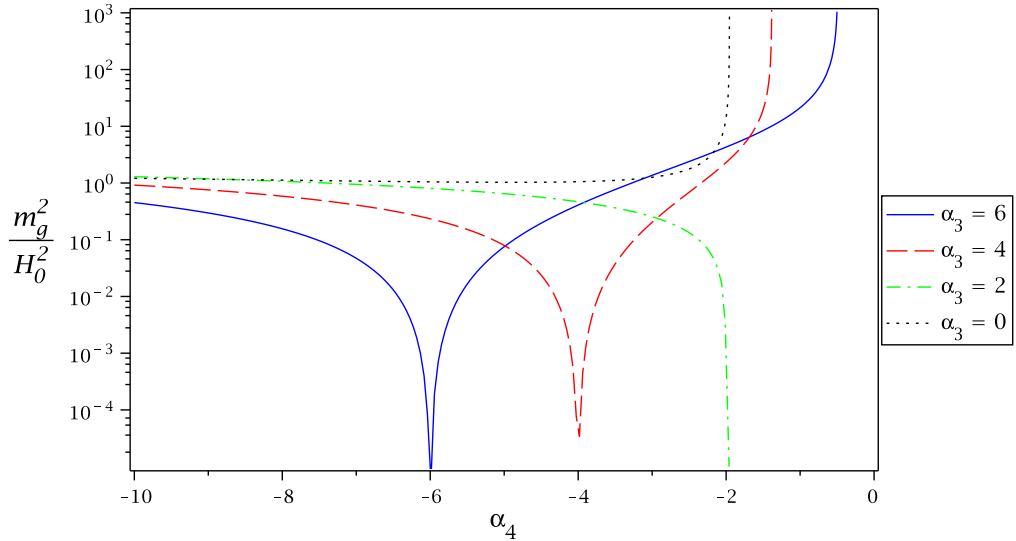, scale=0.7}
\caption{Some typical mass dependence with $\alpha_4$ for some specific values of $\alpha_3$.}\label{fig2}
\end{center}
\end{figure}

This model also gives an expression to the graviton mass depending on the $\alpha_3$ and $\alpha_4$ parameters through $\beta_\pm$:
\be
m_g^2={\Omega_{g}\over \beta_\pm}H_0^2\,.\label{masslimit}
\ee
If we fix some of the parameters, we can see how the mass depends on the others. In Figure \ref{fig2} we show a typical mass dependence with $\alpha_4$ when we set the $\alpha_3$ as a constant and we choose to work with $\beta_-$. Such behaviour is also observed for others positive and negative values of $\alpha_3$. It is easy to see that the mass increases and diverges for some specific value of $\alpha_4$, corresponding to the limit $\beta_- \to 0$. In some cases the mass can also abruptly decrease to zero, as shown in the cases $\alpha_3=6$, $\alpha_3=4$ and $\alpha_3=2$. This shows that the graviton mass is strongly dependent on the $\alpha$'s parameters, although in the limit of very negative $\alpha_4$ values the graviton mass goes to $m_g\simeq H_0^{-1}$.

\section{Conclusion}

In this work we have analysed an open FLRW solution of massive gravity which admits accelerated expansion of the universe, in full concordance to observations. The constraints with recent observational $H(z)$ data were found and the best fit values obtained for the cosmological parameters are $(\Omega_m,\, \Omega_g,\, \Omega_K) = (0.242,\, 0.703,\, 0.055)$, in accordance with the $\Lambda$CDM model, and also point to a nearly open curvature $(K<0)$, for which the model is valid. The graviton mass dependence with the constant parameters $\alpha_3$ and $\alpha_4$ of the model were also analysed, and we have verified a strong dependence with such parameters. We have also obtained that the condition $m_g\simeq H_0^{-1}$ seems dominant for a long range of the parameters $\alpha_3$ and $\alpha_4$, although cosmological observations cannot be used to determine such parameters.

\begin{acknowledgments}
SHP is grateful to CNPq, process number 477872/2010-7. E.L.M thanks CNPq 449806/2014-6. APSS thanks CAPES - Coordena\c{c}\~{a}o de Aperfei\c{c}oamento de Pessoal de N\'{i}vel Superior, for financial support. JFJ is grateful to Unesp - C\^ampus de Guaratinguet\'a for hospitality and facilities. 
\end{acknowledgments}



\begin{thebibliography}{99}

\bibitem{SN} A. G. Riess {\it et al.}, \emph{Observational Evidence from Supernovae for an Accelerating Universe and a Cosmological Constant}, \emph{Astron. J.} {\bf 116}, (1998) 1009; 

S. Perlmutter {\it et al.}, \emph{Measurements of $\Omega$  and $Lambda$ from 42 High-Redshift Supernovae}, \emph{Astrophys.  J.} {\bf 517}, (1999) 565; 

P. Astier {\it et al.}, \emph{The Supernova Legacy Survey: measurement of {$\Omega$}$_{M}$, {$\Omega$}$_{\Lambda}$ and w from the first year data set}, \emph{Astron. Astrophys.}  {\bf 447}, (2006) 31; 

A. G. Riess {\it et al.}, \emph{New Hubble Space Telescope Discoveries of Type Ia Supernovae at $z > 1$: Narrowing Constraints on the Early Behavior of Dark Energy}, \emph{Astrophys. J.} {\bf 659}, (2007) 98.

\bibitem{union2} R. Amanullah \emph{et al.}, \emph{Astrophys. J.} {\bf 716}, (2010) 712.

\bibitem{WMAP} E. Komatsu \emph{et al.}, \emph{Seven-Year Wilkinson Microwave Anisotropy Probe (WMAP) Observations: Cosmological Interpretation}, \emph{Astrophys. J. Suppl.} {\bf 192}, (2011) 18;

D. Larson et. al. , \emph{Seven-Year Wilkinson Microwave Anisotropy Probe (WMAP *) Observations: Power Spectra and WMAP-Derived Parameters}, \emph{Astrophys. J. Suppl.} {\bf 192}, (2011) 16.

\bibitem{planck} P.~A.~R.~Ade {\it et al.}, \emph{Planck 2013 results. XVI. Cosmological parameters},
  \emph{Astron. Astrophys.}  {\bf 571}, (2014) A16 [arXiv:1303.5076 [astro-ph.CO]].
  
\bibitem{farooq} O. Farooq and B. Ratra, \emph{Hubble Parameter Measurement Constraints on the Cosmological Deceleration-Acceleration Transition Redshift}, \emph{Astrophys. J.} {\bf 766}, (2013) L7 [arXiv:1301.5243].

\bibitem{sharov} G.~S.~Sharov and E.~G.~Vorontsova, \emph{Parameters of Cosmological Models and Recent Astronomical Observations}, \emph{JCAP} {\bf 10},  (2014) 57 [arXiv:1407.5405 [gr-qc]].

\bibitem{CC} T. Padmanabhan, \emph{Cosmological Constant - the Weight of the Vacuum}, \emph{Phys. Rep.} {\bf 380}, (2003) 235; 

S. Weinberg,\emph{The Cosmological Constant Problem}, \emph{Rev. Mod. Phys.} {\bf 69}, (1989) 1.

\bibitem{fierz} M. Fierz and W. Pauli, \emph{On Relativistic Wave Equations for Particles of Arbitrary Spin in an Electromagnetic Field}, \emph{Proc. R. Soc. A} {\bf 173}, (1939) 211.


\bibitem{deser} D. G. Boulware and S. Deser, \emph{Can Gravitation Have a Finite Range? }, \emph{Phys. Rev. D} {\bf 6}, (1972) 3368.


\bibitem{hassan1} S. F. Hassan and R. A. Rosen, \emph{Resolving the Ghost Problem in Nonlinear Massive Gravity}, \emph{Phys. Rev. Lett.} {\bf 108}, (2012) 041101; 

S. F. Hassan and R. A. Rosen, {\emph Confirmation of the Secondary Constraint and Absence of Ghost in Massive Gravity and Bimetric Gravity}, {\emph JHEP} {\bf 04}, (2012) 123.

\bibitem{hassan2} S. F. Hassan and R. A. Rosen, {\emph On Non-linear Actions for Massive Gravity}, {\emph JHEP} {\bf 07}, (2011) 009;

S. F. Hassan, R. A. Rosen and A. Schmidt-May, {\emph Ghost-free Massive Gravity With a General Reference Metric}, {\emph JHEP} {\bf 02}, (2012) 026.

\bibitem{hassan3} S. F. Hassan, A. Schmidt-May, and M. von Strauss, \emph{Proof of Consistency of Nonlinear Massive Gravity in the St\"{u}ckelberg Formulation}, \emph{Phys. Lett. B} {\bf 715}, (2012) 335;

\bibitem{rham} C. de Rham and G. Gabadadze, \emph{Generalization of the Fierz-Pauli Action}, \emph{Phys. Rev. D} {\bf 82}, (2010) 044020;

C. de Rham, G. Gabadadze, and A. J. Tolley, \emph{Resummation of Massive Gravity}, \emph{Phys. Rev. Lett.} {\bf 106}, (2011) 231101;

C. de Rham, G. Gabadadze, and A. Tolley, \emph{Helicity Decomposition of Ghost-free Massive Gravity}, \emph{JHEP} {\bf 11}, (2011) 093; 

C. de Rham, G. Gabadadze, and A. Tolley, \emph{Ghost Free Massive Gravity in the Stückelberg Language}, \emph{Phys. Lett. B} {\bf 711}, (2012) 190.

\bibitem{hinterRMP} K. Hinterbichler, \emph{Theoretical Aspects of Massive Gravity}  \emph{Rev. Mod. Phys.} {\bf 84}, (2012) 671.

\bibitem{massiveG} N. Arkani-Hamed, H. Georgi, and M. D. Schwartz, \emph{Effective Field Theory for Massive Gravitons and Gravity in Theory Space}, \emph{Ann. Phys. (N.Y.)} {\bf 305}, (2003) 96; 

  

K. Hinterbichler and R. A. Rosen, \emph{Interacting Spin-2 Fields}, \emph{JHEP} {\bf 07}, (2012) 047;

M. Mirbabayi, \emph{A Proof Of Ghost Freedom In de Rham-Gabadadze-Tolley Massive Gravity}, \emph{Phys. Rev. D} {\bf 86}, (2012) 084006.

\bibitem{volkov} M. S. Volkov, \emph{Exact Self-Accelerating Cosmologies in Ghost-free Bigravity and Massive Gravity}, \emph{Phys. Rev. D} {\bf 86}, (2012) 061502;

A. H. Chamseddine and M. S. Volkov \emph{Cosmological Solutions with Massive Gravitons}, \emph{Phys. Lett. B} {\bf 704}, (2011) 652;

M. S. Volkov, \emph{Stability of Minkowski Space in Ghost-Free Massive Gravity Theory} \emph{Phys. Rev. D} {\bf 90} (2014) 024028.

\bibitem{koba} T. Kobayashi, M. Siino, M. Yamaguchi and D. Yoshida, \emph{New Cosmological Solutions in Massive Gravity} \emph{Nucl. Phys. Proc. Suppl.} {\bf 246}, (2014) 76.

\bibitem{gumru1} A. E. G\"{u}mr\"{u}k\c{c}\"{u}o\v{g}lu, C. Lin, and S. Mukohyama, \emph{Anisotropic Friedmann-Robertson-Walker Universe from Nonlinear Massive Gravity}, \emph{Phys. Lett. B} {\bf 717}, (2012) 295; 

A. De Felice, A. E. G\"{u}mr\"{u}k\c{c}\"{u}o\v{g}lu, and S. Mukohyama, \emph{Massive Gravity: Nonlinear Instability of the Homogeneous and Isotropic Universe}, \emph{Phys. Rev. Lett.} {\bf 109}, (2012) 171101.

\bibitem{gumru2} A. E. G\"{u}mr\"{u}k\c{c}\"{u}o\v{g}lu, C. Lin, and S. Mukohyama, \emph{Cosmological Perturbations of Self-Accelerating Universe in Nonlinear Massive Gravity}, \emph{JCAP} {\bf 11}, (2011) 030;

A. De Felice, A. E. G\"{u}mr\"{u}k\c{c}\"{u}o\v{g}lu, C. Lin, and S. Mukohyama, \emph{On the Cosmology of Massive Gravity}, \emph{Class. Quant. Grav.} {\bf 30}, (2013) 184004;

A. De Felice, A. E. G\"{u}mr\"{u}k\c{c}\"{u}o\v{g}lu, C. Lin, and S. Mukohyama, {\emph Nonlinear Stability of Cosmological Solutions in Massive Gravity}, \emph{JCAP} {\bf 05}, (2013) 035.

\bibitem{rham2} C. de Rham, M. Fasiello and A. J. Tolley, \emph{Stable FLRW Solutions in Generalized Massive Gravity}, \emph{Int. J. Mod. Phys. D} {\bf 23}, (2014) 1443006.

\bibitem{dubowski} S. L. Dubovsky, \emph{Phases of Massive Gravity}, \emph{JHEP} {\bf 10}, (2004) 076.

\bibitem{matas} C. de Rham, A. Matas and A. J. Tolley, \emph{New Kinetic Interactions for Massive Gravity?}, \emph{Class. Quant. Grav.}  {\bf 31}, (2014) 165004 [arXiv:1311.6485v2].


\bibitem{ref05} P. Creminelli, A. Nicolis, M. Papucci and E. Trincherini, {\emph Ghosts in Massive Gravity}, {\emph JHEP} {\bf 09}, (2005) 003.


\bibitem{saridakis} R. Gannouji, Md. W. Hossain, M. Sami, Emmanuel N. Saridakis, {\emph Quasi-dilaton non-linear massive gravity: Investigations of background cosmological dynamics}, {\emph Phys. Rev.} {\bf D 87}, (2013) 123536.

\bibitem{pdg} K.~A.~Olive {\it et al.}  [Particle Data Group Collaboration], \emph{Review of Particle Physics}, \emph{Chin. Phys. C} {\bf 38}, (2014) 090001.



\end{thebibliography}
\end{document}